\documentstyle[twoside,12pt]{article}

\begin{document}
\setlength{\baselineskip}{3.25ex}
\def\maybepagebreak{\vfill\eject}
\parskip 12pt plus 1pt

\def\doublespace {\smallskipamount=6pt plus2pt minus2pt
                  \medskipamount=12pt plus4pt minus4pt
                  \bigskipamount=24pt plus8pt minus8pt
                  \normalbaselineskip=24pt plus0pt minus0pt
                  \normallineskip=2pt
                  \normallineskiplimit=0pt
                  \jot=6pt
                  {\def\smallskip {\vskip\smallskipamount}}
                  {\def\medskip   {\vskip\medskipamount}}
                  {\def\bigskip   {\vskip\bigskipamount}}
                  {\setbox\strutbox=\hbox{\vrule 
                    height17.0pt depth7.0pt width 0pt}}
                  \parskip 12.0pt
                  \normalbaselines}
\def\unetdemi    {\smallskipamount=6pt plus2pt minus2pt
                  \medskipamount=12pt plus4pt minus4pt
                  \bigskipamount=24pt plus8pt minus8pt
                  \normalbaselineskip=16pt plus0pt minus0pt
                  \normallineskip=2pt
                  \normallineskiplimit=0pt
                  \jot=6pt
                  {\def\smallskip {\vskip\smallskipamount}}
                  {\def\medskip   {\vskip\medskipamount}}
                  {\def\bigskip   {\vskip\bigskipamount}}
                  {\setbox\strutbox=\hbox{\vrule 
                    height17.0pt depth7.0pt width 0pt}}
                  \parskip 12.0pt
                  \normalbaselines}
\def\singlespace {\smallskipamount=3pt plus1pt minus1pt
                  \medskipamount=6pt plus2pt minus2pt
                  \bigskipamount=12pt plus4pt minus4pt
                  \normalbaselineskip=12pt plus0pt minus0pt
                  \normallineskip=1pt
                  \normallineskiplimit=0pt
                  \jot=3pt
                  {\def\smallskip {\vskip\smallskipamount}}
                  {\def\medskip   {\vskip\medskipamount}}
                  {\def\bigskip   {\vskip\bigskipamount}}
                  {\setbox\strutbox=\hbox{\vrule 
                    height8.5pt depth3.5pt width 0pt}}
                  \parskip 6pt
                  \normalbaselines}
%


\setcounter{page}{0}
\title{\centerline{A New Ontological View}
\centerline{of the Quantum Measurement Problem}}

\bigskip
\bigskip

\author{
Xiaolei Zhang\\
Code 7218\\
US Naval Research Laboratory \\
4555 Overlook Ave. SW\\
Washington, DC 20375, USA\\
e-mail: xzhang@nrl.navy.mil\\
\\
\\
\\
\\
\\
\\
\\
Naval Research Lab Memoradum Report \\
NRL/MR/7218--05-8883
}

\maketitle

\bigskip
\bigskip
\bigskip

\setcounter{page}{1}
\section*{Abstract}

A new ontological view of the quantum measurement processes is given, 
which has bearings on many broader issues in the foundations of 
quantum mechanics as well.  In this scenario a quantum measurement is 
a non-equilibrium phase transition in a ``resonant cavity'' formed 
by the entire physical universe including all of its material and 
energy content.  A quantum measurement involves the energy and matter 
exchange among not only the system being measured and the measuring 
apparatus but also the global environment of the universe resonant
cavity, which together constrain the nature of the phase transition.
                                                                                
Strict realism, including strict energy and angular momentum conservation, 
is recovered in this view of the quantum measurement process beyond 
the limit set by the uncertainty relations, which are themselves derived 
from the exact commutation relations for quantum conjugate variables.  
Both the amplitude and the phase of the quantum mechanical wavefunction 
acquire substantial meanings in the new ontology, and the probabilistic 
element is removed from the foundations of quantum mechanics, its apparent
presence in the quantum measurement being solely a result of the
sensitive dependence on initial/boundary conditions of the phase
transitions of a many degree-of-freedom system which is effectively
the whole universe.  Vacuum fluctuations are viewed as
the ``left over'' fluctuations after forming the whole numbers of
nonequilibrium resonant modes in the universe cavity.
                                                                                
This new view on the quantum processes helps to clarify many
puzzles in the foundations of quantum mechanics, such as wave-particle
duality, Schrodinger's Cat paradox, first and higher order coherence
of photons and atoms, virtual particles, the existence of commutation 
relations and quantized behavior, etc.  It naturally explains also the
appearance of a hierarchy of structures in the physical universe
as the result of successive spontaneous phase transitions under 
natural boundary conditions, whose occurrence does not need the 
presence of conscious observers to ``collapse the wavefunction''.  
Implications of the new view on the currently successful approaches in 
quantum field theories, such as the renormalization procedure
and Feynman diagrammatic approach, are also discussed.
                                                                                
\section{Introduction}

The advent of quantum mechanics eight decades ago and the accompanying 
Copenhagen interpretation marked the beginning of a new phase of physics 
research in the microscopic domain, whereupon ontological realism was generally 
abandoned for computational proficiency.  Despite the successful demonstration 
of statistical correspondence between quantum theory and experiments
(Wheeler \& Zurek eds. 1983 and the references therein), as well as the 
even more spectacular demonstration of the quantitative agreement 
between the renormalized quantum electrodynamics calculations 
and experimental results on fundamental quantities such as Lamb shift 
and the anomalous magnetic moment of the electron (see, e.g.,
Nambu 1949 and the references therein), a cloud of unease 
lingers over the loss of a concrete and realistic image of the physical 
world, compared to that of the classical world which we were used to.
 
Central to this unease is the quantum measurement problem 
and the associated probabilistic interpretation of the quantum mechanical
wavefunction which predicts in the statistical sense the outcome of
any quantum measurement.  As eloquently summarized in a 
recent article by DeWiit (2005): ``If one accepts the view that 
formalism and reality are isomorphic, then in the quantum theory 
one is obliged to accept a stupendous number of simultaneous realities, 
namely, all the possible outcomes of quantum measurements as well as 
all the possible `classical' worlds that emerge spontaneously from the 
wavefunction of the universe through the phenomenon of decoherence''.
Even without being a whole-hearted convert to the above
many-world view, one is still left wondering about
the underlying implications of the probability interpretation.

As is well-known in the classical world, the statistical and 
probabilistic phenomenology of a physical process does not 
necessarily imply an underlying propabilistic ontology.
Likewise, the propabilistic outcome of the quantum measurement 
does not necessarily warrant a ``many world'' ontology, especially 
if this implies giving up energy conservation which is one of the 
cornerstones of the physical sciences.

The unease results also from the need to artificially draw a line between
the behaviors of the quantum and the classical world.  Since
in the physical universe the large and the small scale phenomena
are often intermingled, and even in quantum measurement of microscopic 
processes the apparatus involved are of macroscopic dimensions, 
this lack of a clear dividing line between the two worlds 
necessarily leads to paradoxes such as ``Schrodinger's Cat
paradox'' (Schrodinger 1935), and the seemingly ridiculous 
need for a conscious observer to ``collapse the wavefunction'' 
in order for the measurement result to settle onto one of 
the many possibilities dictated by quantum calculations
(Von Neumann 1932).

Part of the goals of the current paper is to restore realism 
to the ontological foundations of quantum mechanics, and to bridge 
the connection between the quantum and the classical worlds.  
This work goes beyond the orthodox Copenhagen Interpretation,
and discredits the uncertainty relation as a fundamental law
of physics, and shows rather that the commutation relations,
from which the uncertainty relations can be derived, describe
quantitatively the behavior of the quantized resonant modes in
the universe cavity.  In the new ontology the concepts of 
nonequilibrium phase transition as well as the generalized 
Mach's principle underlie the explanation of  all quantized phenomena.  
A coherent picture can be arrived at which synthesizes also the
results from the past decades on S-matrix theories, 
quantum electrodynamics and gauge field theories, and offers
a clearer explanation of phenomena both in the quantum
as well as in the classical world. 

The organization of the paper is as follows: Section 2
outlines the essential features of the quantum measurement
process, and ascertain that the so-called ``quantum wavefunction 
collapse'' is real and is prevalent in quantum measurements.
It also review several of the previously proposed theories on 
quantum measurement. 
Section 3, which forms the core of this paper, presents the new 
ontological view of the quantum measurement process, gives its 
motivation and main assumptions, as well as the
implications of the new ontology on resolving and coherently
interpreting many issues in the foundations and frontiers
of quantum mechanics.  The remaining unanswered
questions which come into sharper focus under this new picture will
also be outlined. Finally, section 4 summarizes the
conclusions of this work.

\section{The Quantum Measurement Problem}

\subsection{Distinguishing Features of Quantum Wavefunction Collapse}

Whereas a classical measurement usually leads to
a definitive result given by deterministic classical calculations,
the result of a quantum measurement in general can only be predicted 
in a probabilistic sense.  The quantum mechanical wavefunction is 
the most quantitative result from a quantum calculation, and
one can hope for a definitive answer only if the measurement is
towards an eigenvalue of a system already settled onto its
corresponding eigenstate.  If this is not so, the underlying
system is assumed to subsequently ``collapse'' onto an eigenstate 
of the measurement operator with the probability for the choice of
state given by the absolute square of the wavefunction.

First of all, we must convince ourselves that there is sufficient
evidence to indicate that such ``wavefunction collapse'' does
happen during quantum measurement, and that a quantum system
under measurement behaves in essentially different ways from
classical systems.

Most of us are familiar with the position-momentum uncertainty
relation. The quantum mechanical wavefunction of a particle
in momentum eigenstate $\bf{p_0}$ is expressed as the following
plane wave 
\begin{equation}
u_{{\bf{p_0}}} ({\bf{r}})= {1 \over {\sqrt {(2 \pi \hbar)^3} }} 
\exp^{{ i \over \hbar}
{\bf{p_0}} \cdot {\bf{r}}} 
,
\label{eq:1}
\end{equation}
so a momentum measurement will lead to a definite result $\bf{p_0}$.  However,
if this momentum measurement is followed by a positional measurement,
the resulting position can take any value with equal probability, as indicated
by the constant amplitude of (\ref{eq:1}) with respect to the values of
$\bf{r}$. But once the position measurement is performed and obtained a 
specific value $\bf{r_0}$, the wavefunction ``jumps'' to the new form of
\begin{equation}
u_{\bf{r_0}} (\bf{r})= \delta (\bf{r} - \bf{r_0})
,
\end{equation}
the subsequent momentum
measurement once again becomes uncertain, and every value of the momentum
is equally likely to be obtained.  The ``particle'' thus
alternates between possessing a position-eigenstate wavefunction and
a momentum-eigenstate wavefunction while undergoing alternating position 
and momentum measurements.

Other evidence of the wavefunction collapse are found in the experiments
described in Chiao et al. (1994 and the references therein) using photons,
and Berman ed. (1997 and the references therein) using atoms.
In such experiments the photons and atoms are spread out, i.e., 
in mostly momentum eigenstate, just prior to
detection, and are localized at the instant of detection.  This feature
is confirmed in many of the ``delayed choice'' type of experiments 
(Wheeler 1978; Alley et al. 1987; Hellmuth et al. 1987; Briegel et a. 1997). 

Other examples which display the distinctive characteristics of the
quantum measurement process include the Aspect et al. experiment 
(Aspect 1976; Aspect, Grangier \& Roger 1981)
to test Bell's inequality (Bell 1965) and quantum entanglement, which
revealed the existence of the action-at-a-distance type of nonlocal 
communication during the quantum measurement process.
Special relativity and Lorentz invariance appear to be violated
in both the entanglement type experiments and in the broader
quantum measurements which involve an instantaneous wavefunction
collapse.

Another characteristic feature of the wavefunction collapse 
is that energy appears not to be conserved, at least not among the 
measuring apparatus and the system being measured.  For example,
the detection of the position of a particle makes its momentum,
and thus energy, uncertain.  Since this uncertain energy
could in principle be infinite, the question comes as to where
the additional energy has arrived from.  Even for resonant interaction,
it was known that ``...  transitions within short times 
occur not only between states which satisfy the condition 
$E + \epsilon = E' + \epsilon'$
($E$ and $E'$ being the energy of the system before and after the
transition, $\epsilon$ and $\epsilon'$ that of the apparatus).
These states are given preference by resonance only after
a long time.  In practice, after a time $\Delta t$, only transitions
by which $|E + \epsilon - E' - \epsilon'| <= h_{bar} / \Delta t$ are 
of importance (Landau and Peierls 1931)''.
The apparent energy non-conservation is also present in processes
involving virtual particles.

\subsection{Existing Quantum Measurement Theories}

The many aspects of the quantum measurement processes and the various
versions of quantum measurement theories can be found in several books and
references on the subject.  See, e.g., Wheeler \& Zurek (1983),
Braginsky \& Khalili (1992), Namiki, Pascazio, \& Nakazato (1997)
and the references therein.

\subsubsection{The Copenhagen Interpretation}

The orthodox Copenhagen Interpretation (see, e.g. Bohr 1928) for quantum
processes emerged at the wake of Born's propabilistic interpretation 
of the quantum mechanical wavefunction (Born 1926)
and Heisenberg's uncertainty principle (Heisenberg 1927), 
and incorporates also Bohr's own Principle of Complementarity.  
It emphasizes the need for classical description of quantum measurement 
results, but requires a clear (but somewhat arbitrary)
distinction between the quantumness of the system 
being measured, and the classicality of the measuring instrument.
It introduces formally the element of ``wavefunction collapse'' into
the description of quantum measurement, but gives no explanation of
the nature or cause of the collapse.  Neither does it allow any
substantial interpretation of the wavefunction itself.  It is at best only
a compromise for the lack of a true ontological theory.  Its success lies 
solely in the phenomenological description of the quantum world and 
in predicting the quantum measurement results in a probabilistic sense.  
The Copenhagen Interpretation started the trend of ``shut-up and calculate'' 
practice which still dominates the scene of quantum research 
eight decades after its invention.

\subsubsection{``Many World'' Interpretation}

The ``many world'' theory of Everett III (1957) and DeWitt (1970)
did away with the quantum jumps of the standard Copenhagen picture, 
and proposes instead that the quantum world is governed only by the 
linear and continuous evolution of Schrodinger type.  The many possibilities
of reality offered by the Schrodinger's equation are all realized,
and they branch away from one another at every instant.  We are only
aware of one of the possibilities because we reside only on one of
the branches. 

This theory itself obviously is not falsifiable, and does not constitute
a bona fide scientific theory in the sense of Popper (1935).  It also 
conflicts with the realist tradition, and does not offer an explanation of 
where the energy of all the ``branching aways'' comes from.  Neither does it
help understanding quantum features such as the wave-particle duality
in a double-slit type of experiment.

\subsubsection{``Decoherence'' Theories}

The modern decoherence theories (Wigner 1963; Zurek 2002 and the references 
therein) originated from J. Von Neumann's quantum measurement theory 
(Von Neumann 1932).  A common theme in these theories is that the measurement 
process is viewed as a multi-step process, with the initial interaction
of the system being measured and the measuring apparatus 
bringing about a correlation between the two, and their states
becoming entangled. Subsequently, the density matrix
describing such a entangled pure state loses its off-diagonal
elements (either spontaneously as in the case of von Newmann's version,
or through interaction with the environment as in the Zurek's
version), a process which is termed ``decoherence'' or ``reduction
of wavefunction'', and the density matrix of the measurement process
thus becomes that of a mixture state instead of a pure state.
Following this step, one usually has to resort to the intervention
of conscious observers to finally pick from one of the many classical
probabilities indicated by the nonvanishing diagonal elements
of the ``decohered'' density matrix.

The decoherence theory gives no true cause of the
loss of the off-diagonal elements (coupling with a thermal dissipative
environment does not describe the nature of all quantum measurement processes,
not even the most classical ones such as successive position/momentum
measurements).  The further need for conscious observers to intervene
to decide on one of the classical possibilities is also in conflict
with what we observe in the natural world where processes happen
without the intervention of human intelligence.
There is also no explanation in this picture of what
the rest of the potential classical outcomes mean and what happen
to them when a conscious being picks out one among the many.

As we will see below, in the quantum to classical transition, the real 
issue is not the vanishing of the off-diagonal interference term in
the density matrix (i.e.  the so-called decoherence), but rather the 
discontinuous evolution of the wavefunction at the moment of measurement, 
described as a non-equilibrium phase transition.

\section{The New Ontology}

In the classical world, all quantized phenomena we know 
of are related to resonances in a particular type of cavity, either
closed (as in the equilibrium resonance phenomena) or open (as in the 
nonequilibrium resonance phenomena). In particular, the 
nonequilibrium resonance phenomena 
are closely related to the nonequilibrium phase transitions, and the 
structures so formed in the underlying open, far-from-equilibrium systems
are termed ``dissipative structures'' (Glansdorf \& Prigogine 1971; 
Nicolis \& Prigogine 1977).  Some classical examples of dissipative
structures are the turbulent convection cells in atmosphere circulation 
(Benard's instability, see, e.g. Koschmieder 1993 and the
references therein), and the spiral structure in galaxies 
(Zhang 1996,1998,1999,2003,2004).  A close analogy
to such a nonequilibrium phase transition is the three-dimensional stereo
image formation by the human eyes and brain when observing two
separate stereo images: The formation of the three-dimensional
image in the brain is obviously not
a simple linear and sequential optical interaction, but rather 
pattern formation through spontaneous symmetry breaking.  Once formed,
the new pattern is stable against small perturbations.  
Dissipative structures are in dynamical equilibrium - their maintenance
depends on a constant flux of energy and entropy through the system,
and the exchange with their environment.

The universe is an open and evolving system.  If we insist on a
realist interpretation of the physical processes,
as well as on a continuous classical and quantum interface,
we have to seek the origin of the quantized behavior
in resonance phenomena.  Since physical constants
such as Planck's constant are universal, and different types
of interaction processes in vastly different environments
give the same elementary particle properties, such resonant behavior
are expected to originate from the entire matter and energy content
of the universe.  Such a view is closely related to Mach's principle
for explaining the origin of mass and inertia (see, e.g.,
Barbour \& Pfister 1995 and the references therein),
and forms the starting point of the new ontological view of
quantum mechanics we are presenting in this paper.

In what follows in this section, we will first list the
central postulates of this ontology, and then provide
further justifications through the comparison with empirical
evidence, the work of the past few decades in quantum
field theories, and some earlier speculations of the pioneering
workers.

\subsection{Fundamental Postulates}

\begin{itemize}
\item It is assumed that a generalized form of ``Mach's Principle''
governs the operation of the physical universe, and that the local
properties of matter, including the values of fundamental constants
and the forms of physical laws, are determined by the global distribution
of all the matter and energy content in the universe.
\item The physical universe is organized into hierarchies (as reflected
in part in the spontaneous breaking of gauge symmetry in the generation 
of physical laws).  The quantum/classical hierarchy is divided
through spontaneous nonequilibrium phase transitions which form the
macroscopic structures.  These macroscopic structures are
capable of resisting the diffusion/smearing tendency of
pure quantum states governed by Schrodinger-type wave equations
and can thus remain quasi-stable.
\item The quantized nature of fundamental processes originates
from nonequilibrium phase transitions in the universe 
resonant cavity.  The usual quantization procedure by enforcing
the commutation relations is equivalent to establishing modal
closure relations in the universe cavity.  Uncertainty relations 
are only the phenomenological derivatives of
the corresponding commutation relations and they themselves
have no independent fundamental significance.
\item A quantum measurement process is equivalent to
setting up an appropriate boundary condition so that
a phase transition in the joint system of object, measuring 
instrument and the rest of the universe is induced.
\item The quantum mechanical wavefunction describes the substantial
distribution of the underlying matter of specific modal type.
Its absolute square gives the probability for obtaining a particular
configuration in the measurement phase transition,
and its phase encodes the influence of the environment
which determines its subsequent evolution.  The probabilistic element 
is thus removed from the ontology of quantum mechanics.
\item The evolution of physical systems in the universe is 
described by successive stages of continuous wavefunction evolution
and discontinuous phase transitions.  The universe as a whole is
neither described by a single autonomous wavefunction, nor is it
a statistical mixture.  It is organized rather as a hierarchy of
nearly independently-evolving subsystems, with the division of
the hierarchies accomplished by successive stages of phase transitions,
though the interaction and exchanges among the hierarchies also
happen during the same phase transitions.
\item In this picture the vacuum fluctuations are the ``residuals''
of the making of the ``whole'' numbers of non-equilibrium
quasi-stationary modes in the open, nonequilibrium universe cavity.
\end{itemize}

\subsection{Implications on the Foundations of Quantum Mechanics}

The new ontology has immediate implications for obtaining a coherent
picture of the observed quantum phenomena and for bridging the connection
with the classical world.  The effectiveness of these interpretations 
also serves as empirical support for the validity of our fundamental postulates.

\subsubsection{The Meaning of the Quantum Mechanical Wavefunction} 

When Schrodinger first derived his wave mechanics formulation,
he envisioned the wavefunction as representing the electron density,
and a point electron as a superpositional wave packet (Schrodinger 1926a).  
This view was criticized (notably by Lorentz, see, e.g. Moore
1989, p. 216) and Schrodinger himself 
subsequently abandoned it.  There are mainly two problems with this view.
First of all, a wavepacket initially localized in space
is found to disperse with time when the time-dependent Schrodinger
equation is solved, especially if the size of the wavepacket is
not much larger than the de Broglie wavelength. Secondly,
as Schrodinger has commented: ``The $\Psi$ function itself cannot and may not be
interpreted directly in terms of three-dimensional space -- however
much the one electron problem tends to mislead us on this point --
because it is in general a function in (3n-dimensional) configuration
space, not real space'' (Schrodinger 1926).  

The commonly accepted interpretation of the wavefunction in the 
nonrelativistic quantum mechanics is the probabilistic interpretation 
first suggested by Max Born (1926).    Quantum mechanics yields 
only probabilities for the outcomes of atomic events, but the 
probabilities themselves evolve in a precise, deterministic fashion in 
accordance with the Schrodinger equation.  The wavefunction is thus
stripped of any substantial meaning and is considered only as
a probability wave.   In quantum field theories, 
however, a certain realist element can be discerned within the overall 
probabilistic framework.  This is because quantum field theories 
already incorporated many of the modal features implicitly.  

In our current ontology, the realist view of the wavefunction is fully
recovered.  The probability wave is now a physical entity, the chance
factor only enters later, at the moment of measurement.
The fact that the wavefunction exists in the 3n-dimensional
configuration space no longer poses a problem, since this signifies only
the interchange and interrelation of the different parcels of the
modal content among a multi-particle state.  
Thus, the coordinates of choice for the multi-particle
wavefunction could be any of the abstract canonical coordinates of Hamilton's
mechanics, since these can equally effectively express modal
characteristics.  Furthermore, the dispersion of the
wavepacket is a natural feature of the superpositional state for a localized
``particle''.  Schrodinger's earlier confusion lies in regarding
the localized particle as a pure resonant state, and he struggled with
preserving the localization of the wavepacket.  In actuality the momentum
eigenstate, which is spread out in space, is often a more ``natural'' state
for a particle resonance, as we will comment further below,
thus an isolated positional state tends to diffuse towards it.

The probability factor in the result of quantum measurement is due
to the intrinsic sensitive dependence on initial/boundary conditions
of the phase transitions happening in a many degree-of-freedom system.
Just as its classical counterpart, the underlying cause of this
probability factor is deterministic though generally not ``determinable'' 
due to the difficulty of tracking down all the causes in an effective 
infinite degree-of-freedom system which contain the evolving wavefunctions 
and quasi-steady states of all the matter in the universe.

Whereas the amplitude of the wavefunction is correlated with the modal 
density in the configuration space, and can thus lead to the probability of a 
particular measurement outcome as in the Born's interpretation, 
the phase of the wavefunction is related to the interaction fields 
in the gauge field formulation. As C.N Yang puts it, ``all fundamental
forces are phase fields (Yang 1987)''.

The Aharonov-Bohm effect (Aharonov \& Bohm 1959) which shows the influence to
the phase in the region where the nominal field strength is absent demonstrate
that it is the relative change in the phase of the wavefunction of the 
electrons produced by the potential that is physically observable.
The change is not produced by any local interaction of the
potential with the electrons.  Rather, it is dictated by a certain
global property of the potential specified by a gauge function. 

In the new ontology the quantum mechanical wavefunction reflects the 
self-organization of the matter contents in the universe into modal form. 
The wavefunction of a quantum observable usually spreads throughout 
in the infinite space, and the interaction between the different 
observables/modes are global in general, which is the reason that 
quantum mechanics is formulated in the Hilbert space, a natural domain
to describe abstract global modal relations.  
A quantum wavefunction continuously evolves and can be superposed
as in the interference effect.  Measurement phase transition introduces 
nonlinearity and makes the linear interference effect permanent.

\subsubsection{Schrodinger, Heisenberg, and Dirac's Formulation
of Non-Relativistic Quantum Mechanics.  Nature of the Stationary States}

Schrodinger's wave equation describes the continuous evolution for 
a system not already on one of the energy stationary states.
Heisenberg's matrix treatment, on the other hand, deals directly with 
transition probabilities between the discreet stationary states.
The two approaches are complementary, and their formal equivalence
can be demonstrated (Schrodinger 1926b) especially through
the transformation theory developed by Dirac (1926,1927,1958).

However, one can demonstrate that something is missing from these
formulations through a simple example, that of
the trace of a single particle in a Wilson cloud chamber (Mott 1929). 
Since Schrodinger's picture predicts the dispersion of the particle
wavepacket during propagation, and Heisenberg's picture describes
the discontinuous jumps between stationary state, neither
can account for the continuous and yet sharply-defined trace
of a particle in the cloud chamber.  In the new ontology, however,
this continuous and thin trace is interpreted as produced by
a series of environmentally induced phase transitions -- i.e. the
interaction of the particle wavepacket with the cloud chamber
gaseous background.  The thin trace is a dynamical balance of the dispersion
tendency of the single particle wavepacket during propagation,
and the wavefunction collapse tendency onto the successive positional
eigenstates.

A quantum operator in general probes global instead of local 
characteristics.  The operators that commute have the same eigenfunction set.
The eigenvalues and eigenfunctions of operators are related through:
\begin{equation}
A \Psi_{A,n} = a_n \Psi_{A, n}
\end{equation}
The underlying meaning of this operation in the new ontology is that 
certain operators preserve the modal structure of the corresponding
modes. If an operator operates on a state that is not one of its
eigenstates, it induces a phase transition and causes the original state
to collapse onto one of its eigenstates.

It is interesting to note that even though the eigenvalue of a
stationary state is a constant (call it $\alpha$, the eigenfunction 
is in general
time dependent (the particular form of the eigenfunction, if it
is a simultaneous eigenstate of the Hamiltonian as well, is
$|\alpha > \exp ({{-iE_{\alpha} t} \over \hbar})$ (see, e.g., Sakurai 1985).
Therefore quantum mechanical stationary state is a kind of dynamical
equilibrium in constant evolution, consistent with the nonequilibrium
stationary state picture (Zhang 1998) we have proposed.

In practice, apart from stationary states and freely-evolving 
wavefunctions, the density matrix formalism (Von Neumann 1927)
has also been employed to describe quantum systems that are
thought to be statistical mixtures or ensembles.  Since
we adopt a realist ontology, no physical systems will actually
be in a statistical mixture state.  The physical universe is
described as a series of hierarchies defined by successive
phase transitions, which form quasi-stationary structures.
The apparent success of the density matrix approach is understood as the 
intrinsic harmonic nature of the evolution of the parts and parcels of
these quasi-stationary states, and the effect of time average 
in this case mimics the effect of ensemble average.

\subsubsection{Extent of a Quantum Measurement Process}

A quantum measurement is in general a non-local process, and in the 
new ontology its result is not determined soly by the localized measuring 
instrument and the object being measured, as manifested by the probabilistic 
nature of the measurement results.  Quantum measurements also involve
the ``give and take'' with the rest of the universe, as evidenced
in the previously-mentioned position/momentum pair of measurements.
This ``give and take'' with the rest of the universe accounts for the apparent
violation of energy conservation in many quantum measurement processes.
It also provides a natural explanation to the paradoxical fact that the 
accelerated electrons radiate in certain cases (as when they travel freely
in straight lines) and not in other cases (as when they are in the Bohr
atoms).

A pure quantum mechanical resonance remains a modal resonance 
spreading out in space until the moment
of detection.  Detection shows quantum behavior because the 
``remainder'' of the stuff in the universe cavity shows quantum behavior.  
Therefore during the emission or detection of a ``particle''
there has to be a ``give and take'' with the background field
(or the rest of the universe) to make up the difference. 
The detected particle is no longer the same particle 
during propagation because of the exchange with the universal background.

The strongest support for the involvement of the universe resonant
cavity during the quantum measurement process is actually the constancy
of the elementary particle characteristics among many different
physical processes.  Without a global resonant cavity to determine
the modal characteristics, we would not have such things as elementary
particles or for that matter fundamental constants themselves.
The identity of particles is the result of their being the same mode.
Other examples of the extended nature of quantum interactions
are phonons and plasmon-mediated processes in solid state physics.

The global nature of the interaction/phase transition is also reflected in
the ``photon scattering'' experiment of the atom interferometer (Schmiedmayer
et al. 1997). During those single photon scattering events
the phase imprints on the two arms of the atom interferometer
reflected the phase difference
of the photon wavefunction where it intersects the two arms of the
interferometer, even though scattering of a single photon off a single
atom wave is supposedly a single event.  This is no longer so if the two
arm's separation is large enough -- in that case the lost coherence
will not be able to be recovered.

\subsubsection{Wave-Particle Duality}

The wave-particle duality is one of the puzzling characteristics
of the quantum world, which contrasts sharply to the behaviors 
we are used to in the classical world.  The wave-particle duality 
is partly what Bohr based his Complementarity Principle upon (Bohr 1928).  
Modern quantum optical experiments show that a free-propagating beam 
of photons show simultaneously the wave and the particle characteristics
(Aichele et al. 2004).

The wave-particle duality is manifest most clearly in the de Broglie 
relation ($p = h/\lambda$) and Planck relation ($E=h \nu$), 
each on one side indicates a pure wave characteristic ($\nu$ and $\lambda$)
and on the other a pure particle characteristic ($E$ and $p$).
The seeming contradictory characteristic is easily clarified in the
new ontology.  Here a particle is more of a pure resonance
when it is a wave mode and is spread out.  When it is a localized particle,
it is in a mixed resonant state, or the superposition of pure states.

In fact, in the quantum field theories, only the fields are localized,
but field quanta are spatially extended.  These spatially distributed
field quanta arrive from the first approximation of the solution
of field equations in the non-interacting limit, which simplifies
analysis and is the source of the name ``particle''.

\subsubsection{Uncertainty Principle, Quantum Commutation Relation,
and Classical Poisson Bracket}

Heisenberg (1926) wrote in a letter to Pauli explaining the
implications of the position-momentum commutation relation and
the associated uncertainty principle:
``The equation $pq-qp = h/2 \pi i$ thus corresponds always in
the wave representation to the fact that it is impossible
to speak of a monochromatic wave at a fixed point in time
(or in a very short time interval) ... Analogously, it is
impossible to talk of the position of a particle of fixed
velocity''.  Schrodinger (1930) subsequently proved the general
relation between the commutation relations and the uncertainty
relations. The founders of quantum mechanics thus were
well aware of the connection between the two kinds of relations.
A general derivation of the uncertainty relations from the
commutator relations can be found in Sakurai (1985, p.34).

The uncertainty principle is thus just another way of writing 
the commutation relations, which themselves are deterministic.
The uncertainty principle itself seemed to have later acquired more
prominence in the quantum discussions only because of its 
intimate relation to the probabilistic outcome of quantum 
measurements.

It is well-known that quantum commutation relations can be
linked to classical Poisson brackets in the ``correspondence
principle'' sense of the quantum to classical transition (Goldstein
1980, p.399).  This tight correlation between the two formulations 
demonstrates once again that the physics in the two regimes
is closely linked, and the form of classical dynamics itself is already
a result of the modal selection in the universe resonant cavity
(we will comment more on this in section 3.2.10).

The special characteristic of the quantum commutation relations 
of course is the ever-present Planck's constant h, which sets the
scale of quantum interactions.  The historical introduction
of Planck's constant of course is through the well-known
exploration of the functional dependence of blackbody radiation
on wavelength and temperature (see Agassi 1993 for a chronological
account), which leads to Planck's discovery of the famous $E=h \nu$ 
equation signaling the beginning of the quantum era.  
Once the value of h is obtained by a comparison
of Planck's blackbody formula
with the empirical blackbody curve, its subsequent
entry into quantum mechanics is through a series of ``correspondence''
type of analogies: i.e. Bohr's assumption of discrete energy levels of
atoms as the origin of discrete spectra; Heisenberg's ingenious application
of correspondence principle which leads to the first incidence of
a quantum commutation relation; de Broglie's generalization of the
quanta energy and momentum relations to that of material particles;
and Schrodinger's derivation of his famous wave equation through
the integration of de Broglie formula and classical mechanics.
Thus, the root of quantized behavior (or at least our discovery of it)
goes back to the source of cosmic blackbody radiation (even though
at the time of Planck he was mostly focused on the blackbody radiation
in a small box), which we know is related to the universe 
resonant cavity as a whole!

The relation between the quantum and classical modal behavior allows
us to borrow the tools developed for analyzing classical nonequilibrium 
structures (Zhang 1996, 1998) and resonant phenomena (Cohen 1995) towards
a deeper understanding of the behavior of quantum systems.

\subsubsection{Quantum Vacuum}

After quantizing space with a set of modes using the commutation or
anticommutation relations, we expect to end up with some leftovers, 
and these leftover fractional modal content we propose is the 
composition of vacuum fluctuations.

The vacuum field fluctuates because the resonant components keep evolving
in the nonequilibrium universe cavity, just like in another example of
such a nonequilibrium phase transition, that of the spiral structure in
galaxies (Zhang 1998), where the individual star's trajectory keeps evolving 
and moving in and out of the spiral pattern even though total energy is 
conserved and the spiral pattern is meta-stable.

This picture provides a possible explanation of why 
many fundamental physical effects (such as Lamb shifts,
Casimir effects, spontaneous emission, van der Waals forces, and
the fundamental linewidth of a laser) can be explained equally successfully
by adopting either the vacuum-fluctuation point of view
or the source-field point of view (see, e.g., Milonni 1994 and the
references therein).

Addition of metal plates as in Casimir effect changes the boundary
condition of the entire vacuum, and force is thus needed to put the plates in.
The energetically most favorable arrangement of the resonances is thus
changed.  So is the modal structure.

The relation of field quantization and vacuum fluctuation  may also be
related to the ``fluctuation-dissipation theorem''.  The dissipative
leaks into the vacuum is needed to maintain the stability of the nonequilibrium
modes.   The universe also has to be constantly evolving in order for the 
fundamental resonances to be stable.
So the un-saturatedness and the constant evolving nature of the universe
maybe a prerequisite for setting up the fundamental laws as we
observe today.  

\subsubsection{Lifetime of Levels and Virtual Particles}

Since quantizations of physical systems do not happen in a closed box, 
but rather in an open and evolving universe, there is 
a continuous exchange of any given mode with the underlying continuum.  
The modal characteristics thus formed are not always sharp, which allowed
the existence of wavepackets, as well as the finite lifetimes of
``fundamental particles'' and ``eigenstates''.  These quasi-stationary 
states are generally in dynamical equilibrium.  The quantization 
characteristic is manifestly the sharpest during the emission/detection 
process, which is a true non-equilibrium phase transition -- but even
there the emitted photons/particles may not immediately be in a
momentum eigenstate.  A photon during propagation is not always in a \
pure momentum eigenstate, but has smear, as we will comment further below.

Virtual particles are those which appear in a quantum electrodynamic
calculation and do not satisfy energy and angular momentum conservation: 
They are ``not on the mass shell'' and are represented by the
internal lines in Feynman diagrams (Feynman 1949;
Dyson 1949). Their existence is another indication 
that a quantum phase transition involves the rest of the universe to 
``close the loop'', and the conservation relation is restored for resonant 
interactions only when the phase transition is complete. 

\subsubsection{First and Higher Order Coherence of Photons and Atoms.
Identical Particles.  Blackbody Radiation}

When Dirac commented that ``A photon interferes only with itself'' (Dirac 1958),
he referred to the first order coherence properties of photons.
Subsequent intensity interferometry experiments (Hanbury Brown \&
Twiss 1954, 1956; Hanbury Brown 1974)
had revealed that photons do interfere with one another, 
which are the higher order coherence properties of photons. 
Such first and higher order coherence properties were also confirmed
for atoms in the atom interferometry experiments (see the contributions
in Berman eds. 1997).

In the current ontology, the first order coherence of the atoms and
photons reveals their underlying wave and modal nature, whereas the
higher order correlation is a manifestation of the finite-Q nature 
of the universe resonant cavity, resulting in ``non-pure'' spatial modes
which entangles the individual particles or field resonances.
Due to the entanglement (as reflected in the Bose-Einstein or
Fermi-Dirac statistics, for example, which can be regarded as different
types of global mode organizations), after quantized emission 
a photon has a tendency to merge back to the universal ``soup'' 
of the background photon flux during propagation, unless the photon
flux is so low that it can be described as spatially and temperally
separated monophotonic states (Hong, Ou, \& Mandel 1987), in which
case its degree of second order coherence $g^2(0) <1$ as is appropriate
for photons in the non-classical photon number states (Loudon 1983).  
The analytical expressions for the degree of second order coherence 
for bosons and fermions show different expressions according to
their respective wavefunction symmetries (Scully \& Zubairy 1997,
p.125), and these statistics are only meaningful when the particle
flux is high enough.

The existence of second or higher order coherence does not require that 
the wave be non-classical (i.e. quantized, see Hanbury Brown 1974).
In some sense, a classical wave itself embodies the correlation between
its constituent parts, and there is
no clear dividing line between the classical and the quantum worlds.
It is only by convention (or chronology) that we refer to the particle nature 
of the photons as nonclassical and the wave nature as classical, whereas
for massive particles our convention is the reverse.

The correlation and entanglement characteristic is needed to 
make sense of the coherent length and coherent time of photon waves.  
It would be meaningless to talk in such terms if each photon 
is completely independent of the other ones.  For coherent length 
to matter for the easiness of wavefunction collapse during a 
quantum measurement, as in the comparison of
beam splitter decoherence in either the laser or Bose-Einstein 
condensate cases (Zhang 2005), the relevant particles
must have mutual interactions during propagation. Otherwise only
the first order coherence will matter.  
Furthermore, the correlation between the coherence length 
or coherence time with the bandwidth of electromagnetic radiation 
also indicates the global nature of the underlying modes, that the universe 
cavity is connected and shared among the different energy quanta.

The internal correlation is also revealed in the equation of
blackbody radiation.  Ehrenfest (1911) had already realized that
independent quanta could only lead to Wien's law, but not Planck's law.  
Thus, if light quanta were to be described by the Planck distribution,
they had to lack statistical independence and show wave-type correlation
through satisfying Bose-Einstein statistics
(Ehrenfest \& Kamerling-Onnes 1915; Bose 1924).

\subsubsection{Spontaneous Quantum Measurements in Naturally Occuring
Orders}

Under the new ontology there is no longer a dichotomy between 
the classical and the quantum world.  The classical systems consist 
of subunits where ``wavefunction collapse'' have already been induced 
by nature, through naturally occurring boundary conditions.  
A macroscopic object does not possess an overall quantum mechanical
wavefunction that freely evolves.  The different hierarchies of
the macroscopic system established by non-equilibrium phase
transitions have quasi-autonomy.

The spontaneous nature of the phase transitions in natural systems
helps to resolve ``Schrodinger's Cat'' type of paradoxes 
(Schrodinger 1935), since a
naturally occurring ``quantum measurement'' does not have to
involve a conscious observer.  The cat in question was already in
a definitive live or dead state before the observer opened the box,
and not in a linear superposition state of the kind $a \cdot Alive + b \cdot
Dead$.  The possibility that naturally
occurring orders are results of non-equilibrium
phase transitions also explains the stability
and reproducibility of these natural orders, i.e., the result
of the non-equilibrium phase transitions is insensitive to the
{\em details} of the initial-boundary conditions, and depends
only on the gross nature of these conditions.

Such views have also been expressed by the founder of the
dissipative structure theory I. Prigogine (1997):``But once it is
shown that instability breaks time symmetry, the observer is
no longer essential.  In solving the time paradox, we also solved
the quantum paradox''.

\subsubsection{The Form and Origin of Physical Laws.  Values of
Fundamental Constants}

The physical laws, both classical and quantum,
can often be derived from least action or variational
principles (Goldstein 1980; Weinberg 1996). 
This behavior generally reflects the fact that the energy 
content of the process under concern is distributed globally, and the
process samples the environmental/boundary conditions of the entire space 
of relevance.  In the case of the classical trajectory of particles, 
this is reflected in Feynman's rule of sum over all possible paths
(Feynman \& Hibbs 1965).
The quantum laws of count statistics for identical particles 
(from which the Planck's law was derived from)
are valid also because the universe is a connected 
resonant cavity. 

The relations between symmetry and the conservation (Noether's
theorem, see, e.g. Goldstein 1980) indicate that the 
dynamical equations (both classical
and quantum) have already incorporated (or is consistent with)
the symmetry of spacetime, i.e., the equations and laws themselves are 
selected by the universe resonant cavity.  This is reasonable since both
the equations and the fundamental constants are likely to be a result of
spontaneous breaking of gauge symmetry, whose occurrence is closely
related to the state of the matter in the universe.

If values of the fundamental constants are determined
by the characteristics of the universe resonant cavity,
the variation of the values of these ``constants'' with time
(such as the recent observation of the likelihood of the
variation of the fine-structure constant) would be naturally
expected if the universe resonant cavity changes with time,
e.g. due to the expansion of the universe and the associated
evolution of matter distribution.

\subsection{Connections with the Past and Current Practices 
in Quantum Theories}

The discussions in the following had benefited greatly
from the survey studies of the development
of 20th century field theories of Cao (1997).

\subsubsection{S-Matrix Theory}

In the 1950s and 60s the S-matrix theory, first proposed by Heisenberg 
(1943a,b, 1944) and later espoused by Chew and collaborators (Chew 1961 and
the references therein),  was enjoying the popularity 
the quantum field theories later enjoyed.  The motivation for the
theory is from processes such as Compton scattering which shows 
that light quanta comes in and goes off as approximate plane waves.

In this theory, the dynamics were not specified by a detailed
model of interaction in spacetime, but were determined by the
singularity structure of the scattering amplitudes, subject to
the requirement of maximal analyticity.  The success of the 
S-matrix approach is likely to be due to the fact that the 
underlying physics obeys global, modal characteristics.
The calculation scheme in this theory was carried out on the
mass shell, which is fine since it only concerns the input and output
states, or the results of phase transition, and hence involves 
only asymptotic states where particles were outside each other's region
of interaction.

\subsubsection{Quantum Field Theories:
Renormalization and Feynman's Diagrammatic Approach}

In quantum field theories, such as quantum electrodynamics (QED), 
the interaction is transmitted by the discrete quanta of the 
electromagnetic field, and the interacting particles are 
considered the quanta of a fermion field. 
All such quanta can be created and destroyed.

Quantum field theory describes local interactions between
particles and fields.  In view of our previous discussions,
that all interactions in the quantum world should be global
by nature since they are in essence phase transitions in
the universe cavity, we wonder how this could be reconciled
with the apparent success of quantum field theory.  In the 
following, we argue that many of the approaches adopted in
quantum field theory calculations have in fact incorporated
the global nature of the underlying processes.

For many quantum field processes such as scattering,
the amplitudes are calculated from the chosen input and output 
states, which are generally expressed in plane wave forms.
Invariably in these calculations the exact behavior of
particles at the supposed ``locations'' of interaction
are not specified in detail.  This shows that the supposedly
local quantum field calculations incorporated features of
a global approach as in the S-matrix approach (Chew 1961).
Scattering effectively becomes persistent scattering and neither
starts nor stops.

Some of the common practices in quantum field calculations, 
such as Feynman's diagrammatic approach (Feynman 1949; 
Dyson 1949, which has its lineage in the S-matrix approach),
are also integral representations of the entire
phase transition process, described in terms of
input and output states and omitting any detailed
description of the ``on-location'' behavior of
the interaction and particle generation/annihilation.

Another feature that indicates the insufficiency of
a purely local description is the need of renormalization
in quantum field theories.  It was first realized by
Weisskopf (1939) that the electromagnetic behavior of the electron 
in quantum field theory is not completely point-like,
but is extended over a finite region.  This extension, 
keeps the divergence of the self-energy within the logarithmic divergence
which made the renormalization procedure possible.

The elementary phenomena in which divergences occur are the polarization 
of the vacuum and the self-energy of the electron (Schwinger 1948, 1949a,b).  
Each of these phenomena describes the interaction of one field with the vacuum 
fluctuations of the other field -- once again a reflection of the
global nature of quantum interactions.  The overall effect 
is to alter the values of the electron charge e and mass m,
which are compensated by charge and mass renormalization.

The local interactions described by quantum field theories
are for parts and parcels of fields.  Yet like the case of any 
dissipative structures, the local characteristics are determined
by the global organization and the global dynamical equilibrium.
The local organization on the other hand reflects features
of the global organization, in this case through the need of
renormalization to obtain a self-consistent theory.

The need for renormalization of any quantum field theories results
not only from the local/global dichotomy between quantum field
theories and quantum process, it also originate from the
fundamental incompatibility of quantum theory with relativity
theory. QED as a relativistic generalization of quantum
mechanics is not self-consistent (other than in its non-interacting
limit: in which case we obtain pure modes and no non-equilibrium
phase transitions).  The effect of spontaneous
quantum measurements to order the processes into hierarchies
is incorporated through the procedure of renormalization.  Empirical
input is needed to construct a self-consistent theory due to
the singular nature of phase transitions whose details
cannot be models in a top-down deductive type of analysis.
As Gross (1985) puts it: ``Renormalization is an expression of the variation
of the structure of physical interactions with changes in the scale
of the phenomena being probed''.

\subsubsection{Gauge Field Theories, Spontaneous Symmetry Breaking, 
and Effective Field Theories}

Gauge field theories emerged within the framework of quantum
field theories, and is very powerful in exploring global features of 
field theories.  In this theory the phase of a wave function 
becomes a new local variable.  The invariance of a theory under 
the {\em} global phase change, or a gauge transformation
of the first kind, entails charge conservation.  The {\em local} 
gauge invariance is related to the electromagnetic interaction.

The existence of gauge principle for the determination of the
forms of fundamental interactions indicates that the dynamical
laws have been co-selected in the universe resonant cavity
as the processes themselves.  The forms of the fundamental interactions
must preserve and alter the modal structure of the underlying matter
and energy distribution in a globally self-consistent way.
In this sense, the experience gained in obtaining the globally
self-consistent solutions for classical nonequilibrium
structures (Zhang 1998) can be used to gain deeper insight into the
organization structure of quantum laws and fields.

Spontaneous symmetry breaking was developed as a mechanism to
preserve gauge invariance when dealing with massive gauge quanta.
The generation of fundamental interactions through spontaneous 
symmetry breaking is viable since in general ``The solutions of the 
equations will possess a lower symmetry than the equations themselves''  
(Landau 1958, quoted in Brown and Rechenberg 1988).  The spontaneous 
symmetry breaking first proposed by Heisenberg and collaborators 
(Durr et al. 1959) is concerned with the low-energy behavior of the solutions
and asserts that some low energy solutions exhibit less symmetry than
the symmetry exhibited by the Lagrangian of the system, while others
possess the full symmetry of the system.

Since the late 1970s theorists gradually realized that the high-energy 
effects in gauge theories can be calculated without taking the cutoff 
in the renormalization approach to infinity.  The cutoffs in these 
Effective Field Theories (Weinberg 1980) acquire physical significance 
as the embodiment of the hierarchical structure of quantum field theory, 
and as a boundary separating energy regions which are separately 
describable by different sets of parameters and different 
interactions with different symmetries.  Non-renormalizable theories 
can also be treated as effective field theories incorporating
empirical input from the experiments.

Thus it appears the new picture offered by the modern development of field
theories points also to a kind of hierarchical organization of the
high-energy world through successive phase transitions. Each layer of the
hierarchy is quasi-autonomous, each has its own ontology and 
the associated ``fundamental laws''.  The ontology and dynamics of 
each layer are quasi-stable, nearly (but not totally)
immune to whatever happen in other layers.

\subsection{Issues Requiring Further Study}

\subsubsection{Connection of Quantum Mechanical Measurement Process
with Relativity Theories}

The quantum measurement processes, especially the Einstein-Podolsky-Rosen
(1935) type or delayed-choice (Wheeler 1978) type of experiments, 
clearly show both the global extent of the wavefunction 
and the superluminal nature of the wavefunction collapse.

Since special relativity dictates that physical signals travel
with a speed less than the speed of light c, this element of the
wavefunction collapse requires further study.
Furthermore, we could ask questions such as whether the speed of 
a single-particle wavefunction collapse depends on the spatial spread of the
wavefunction, and whether the instantaneous wavefunction collapse
corresponds to instantaneous energy propagation as well.

In fact, the ontology of relativity theories may itself
need to be further elucidated. Questions we could ask include:
Why is there a speed-of-light limit?  Is it only there 
for certain levels of hierarchy and not for others (see next section)?
If in the eye of light, the world is but a single point, how could
changes and evolution ever be possible?  Does that mean for light
the external degrees-of-freedom get transformed into the internal 
degrees-of-freedom to allow evolution?
The essential incompatibility of relativity theory and quantum mechanics,
i.e. the loss of Lorentz invariance in quantum measurement processes
needs to be further explored.
 
\subsubsection{Hierarchy of Laws and Mach's Principle}

We have commented that it appears that the speed of light limit
holds for classical physical processes as well as non-measurement
or non-phase-transition type quantum processes.
This is an indication that the workings of the physical world
may be organized into hierarchies.  Even though in the lower hierarchy
the speed of light is a fundamental limit to signal propagation,
the same is not true for wavefunction collapse.  

Furthermore, we expect that the establishment of the fundamental laws and 
the selection of fundamental constants through the resonant interaction
in the universe cavity according to Mach's principle are also themselves
processes which are not constrained by the speed of light limit:
For otherwise the mere size of the universe and the cycle of 
propagations needed to resonantly establish a fundamental constant
would make it impossible for these constants and laws to be universal
and atemporal (at least over short time span).  
The statistical laws of physics, such as Einstein's A and B coefficients 
for spontaneous and stimulated emission (Einstein 1917), and Planck's 
radiation law, would not be valid and universal if the impact of the entire 
content of the universe is not communicated instantaneously.
The variational approach for the selection
of laws (as indicated by the Lagrangian type of derivation)
only works if all possible paths are ``explored'' acausally. 
The fact that in the variational approach the different path explored
are not constrained by the speed of light limit is evidenced
from Feynman's path integral formulation (Feynman \& Hibbs 1965).

As commented by Barbour \& Pfister (1995),
``It is often not sufficiently appreciated how kind nature
has been in supplying us with `subsystems' of the universe which
possess characteristic properties that can be described and measured
almost without recourse to the rest of the universe. ... On the
other hand, it is evident that basic concepts such as `inertia'
and `centrifugal force' cannot be understood and explained within
the context of the subsystems themselves, but at best by taking
into account the rest of the universe''.  That refers to Mach's
original speculations on the origin of inertia.  By the same token
other properties of the universe cavity such as the quantization
unit h are expected to be determined by the entire universe as well.

Questions requiring further work to answer include: What 
characteristics of the universe resonant cavity generate the
fundamental constants (presumably as eigenvalues)?  How do the
different levels of hierarchy interact? etc.

\subsubsection{Causality of the Quantum Processes}

The global nature of the quantum phase transitions also demands
us to alter our traditional notions of cause and effect.
The boundary between causes and effects are now blurred, and
there is now only an intertwined co-evolution of all the 
parts of the universe resonant cavity, strictly speaking --
though the hierarchical structure of orders in the universe
due to the successive phase transitions allows the emergence 
of macroscopic objects and classical phenomena which conform 
to our notion of the cause-and-effect.

At a deeper level though, we could not ignore the fact that
we have now a kind of more distributed and global causality.
It was already known since the early days of quantum mechanics
that radiative phenomena cannot be analyzed causally.  The stability
of the atom shows that there cannot be a classical type of causal
process in the radiability of the atom.  Scattering of quantum particles
likewise is also a distributed process which could not be analyzed
as a causal sequence.  In this picture, forces as exchange of particles 
serve to realize an energetically more favorable configuration.

Other evidence of the new causality include the description of many
natural processes using analytical signals.  To know an analytical
signal at time t we have to know the real signal at all times: This
once again shows the global influence of the modal behavior.

Mach himself summarizes this intermingled state of cause and effect
in the self-sustained universe as follows: ``Thus the law of causality
is sufficiently characterized by saying that it is the presupposition
of the mutual dependence of phenomena.  Certain idle questions,
for example, whether the cause precedes or is simultaneous with the
effect, then vanish by themselves'' (Mach 1911); as well as
``The universe is not given us twice, but only once''.

The many consequences of this new meaning of quantum causality
need to be further explored.

\subsubsection{The Possibility of Free Will in A Universe Governed
by Deterministic Physical Laws}

Does the deterministic ontology presented in this paper
prohibit the emergence of free-will of sentient beings, which
we know from personal experiences do exist?

From the work of Yang and Lee (1952), we know that the sharpness of
the phase transitions in classical statistical systems is related
to the effective number of degree-of-freedom of the system.
In the new ontology, the effective degree-of-freedom of the physical phenomena
in the universe resonant cavity is non-denumerably infinite, since
we are dealing with a true continuum made up of the smooth reservoir
of energy, and the discreteness we observe in the physical universe
is merely resonant modal behavior formed in the universe cavity.

The infinitely-sensitive dependence on the details of perturbations 
of an infinite degree-of-freedom
system is likely to be the cause of the indeterministic 
factor in the generation of free will, though the detailed workings of
the free-will generation process still require further study.  
This indeterministic factor in free will is not to be confused with 
the quantum uncertainty relations, which are themselves only a derived 
property of the exact quantum commutation relations.
The indeterministic factor in the phase transitions of an infinite
degree-of-freedom system allows chance to play a role, 
and introduces an element of 
history into the course of evolution of the universe.

\section{Conclusions}

In this paper, we argue that even though the quantum mechanical paradigm 
is the correct description of microscopic phenomena, unitary evolution
is not the only guiding principle.  To that we need to add the principle
of phase-transition in the universe resonant cavity, and this principle 
is responsible for organizing both the physical laws and natural orders 
into hierarchies.  Both the amplitude and the phase of the quantum 
mechanical wavefunction are substantial in nature, and the probability
element is removed from the foundations of quantum mechanics.
An effectively classical ontology is provided for the quantum processes.

The new ontology on the foundations of quantum mechanics and especially 
on the nature of quantum measurements leads to a coherent interpretation 
of a wide spectrum of phenomena in the quantum as well as the classical 
world.  It also provides a framework for understanding and coherently 
linking many diverse results in quantum field theories which have been
accumulated over the past few decades.  As a working hypothesis, it could 
serve to stimulate a new synthesis on our understanding of the workings 
of the physical universe.

This research was supported in part by funding from the Office of
Naval Research.

\section{References}

\noindent
Agassi, J. 1993, Radiation Theory and the Quantum Revolution 
(iBasel: Birkhauser)

\noindent
Aharonov, Y., \& Bohm, D. 1959, Phys. Rev. (Ser. 2) 115, 485

\noindent
Aichele, T., Herzog, U., Scholz, M., \& Benson, O. 2004, quant-ph/0410112

\noindent
Alley, C.O., Jakubowicz, O.G., \& Wicks, W.C. 1987, in Proc. Second
International Symposium Foundations of Quantum Mechanics, eds.
M. Namiki et al. (Tokyo: Phy. Soc. Japan), p.36

\noindent
Aspect, A. 1976, Phys. Rev. D14, 1944

\noindent
Aspect, A., Grangier, P., \& Roger, G. 1981, Phys. Rev. Lett. 47, 460

\noindent
Barbour, J. \& Pfister, H. 1995, Mach's Principle: From
Newton's Bucket to Quantum Gravity (Boston: Birkhauser)

\noindent
Bell, J.S. 1965, Physics, 1, 195

\noindent
Berman, P.R. ed. 1997, Atom Interferometry (San Diego: Academic Press)

\noindent
Bohr, N. 1928, Nature, 121, 580

\noindent
Born, M. 1926, Zeitschrift fur Physik, 37, 863, English version in
Wheeler \& Zurek eds. Quantum Theory and Measurement (1983), p. 52

\noindent
Bose, S.N. 1924, Zeitschrift fur Physik, 26, 178

\noindent
Braginsky, V.B.,  \& Khalili, F.Y. 1992, Quantum Measurement,
(Cambridge: Cambridge Univ. Press)

\noindent
Briegel, H.J., Englert, B.G., Scully, M.O., \& Walther, H. 1997,
in Atom Interferometry, ed. P.R. Berman (San Diego: Academic Press), p. 217

\noindent
Brown, L.M., \& Rechenberg, H. 1988, in Landau's work on quantum Field
theory and high energy physics (1930-1961), 
Max Planck Institute preprint, MPI-PAE/Pth 44/88 (July 1988)

\noindent
Cao, T.Y. 1997, Conceptual Developments of 20th Century Field
Theories (Cambridge: Cambridge Univ. Press)

\noindent
Chew, G.F. 1961, S-Matrix Theory of Strong Interactions (New York:
Benjamin)

\noindent
Chiao, R.Y., Kwait, P.G., \& Steinberg, A.M. 1994, in ``Advances in Atomic,
Molecular, and Optical Physics'', B. Bederson \& H. Walther eds., vol 34,
(New York: Academic Press), p. 35

\noindent
Cohen, L. 1995, Time-Frequency Analysis (New Jersey: Prentice Hall)

\noindent
DeWitt, B.S. 1970, Phys. Today, 23, 30

\noindent
DeWitt, B.S. 2005, Physics Today, January issue
 
\noindent
Dirac, P.A.M. 1926, Proc. Royal Soc. London, vol. 112, No. 762, p. 661

\noindent
Dirac, P.A.M. 1927, Proc. Royal Soc. London, vol. A113, p. 621
 
\noindent
Dirac, P.A.M. 1958, The Principles of Quantum Mechanics, 4th edition (Oxford:
Oxford Univ. Press)

\noindent
Durr, H.P., Heisenberg, W., Mitter, H., Schlieder, S., \& Yamazaki, K. 1959,
Zeitschrift fur Naturforschung, 14A, 441

\noindent
Dyson, F. 1949, Phys. Rev. 75, 486

\noindent
Ehrenfest, P. 1911, Ann. Phys., 36, 91

\noindent
Ehrenfest, P. \& Kamerling-Onnes, H. 1915, Proc. Amsterdam Acad. 23, 789

\noindent
Einstein, A. 1917, Phys. Z. 18, 121, English translation by
D. ter Harr 1967, The Old Quantum Theory (New York: Pergamon Press), p.167

\noindent
Einstein, A., Podolsky, B., \& Rosen, N. 1925, Phys. Rev. 47, 777

\noindent
Everett III, H. 1957, Rev. Modern. Phys. 29, 455

\noindent
Feynman, R.P. 1949, Phys. Rev. 76, 769

\noindent
Feynman, R.P., \& Hibbs, A.R. 1965, Quantum Mechanics and Path Integrals
(New York: McGraw-Hill)

\noindent
Glansdorf, P., \& Prigogine, I. 1971, Thermodynamic Theory of
Structure, \indent Stability and Fluctuations (New York: Wiley-Interscience)

\noindent
Goldstein, H. 1980, Classical Mechanics, 2nd ed. (Menlo Park: Addison-Wesley)

\noindent
Gross, D. 1985, in Recent Developments in Quantum Field Theories, J. Ambjorn,
B.J. Durhuus, and J.L. Petersen, eds., (Amsterdam: Elsevier), p. 151

\noindent
Hanbury Brown, R. 1974, The Intensity Interferometer (London: Taylor and 
Francis)

\noindent
Hanbury Brown, R. \& Twiss, R.Q. 1954, Philos. Mag., Ser. 7, No. 45,
663

\noindent
Hanbury Brown, R. \& Twiss, R.Q. 1956, Nature, 178, 1046

\noindent
Heisenberg, W. 1926, in Pauli, W., Wissenschaftlicher 
Briefwechsel mit Bohr, Einstein, Heisenberg, u.  A., 
Band I: 1919-1929, eds. Hermann, A., Meyenn, K.V., and Weisskopf, V.E.
(Berlin:Springer, 1979)

\noindent
Heisenberg, W. 1927, Zeitschrift fur Physik, 43, 172, English version
in Wheeler \& Zurek eds. Quantum Theory and Measurement (1983), p. 62

\noindent
Heisenberg, W. 1943a, Zeitschrift fur Physik, 120, 513

\noindent
Heisenberg, W. 1943b, Zeitschrift fur Physik, 120, 673

\noindent
Heisenberg, W. 1944, Zeitschrift fur Physik, 123, 93

\noindent
Hellmuth, T., Walther, H., Zajonc, A., \& Schkeich, W. 1987, Phys. Rev. A, 35, 2532

\noindent
Hong, C.K., Ou, A.Y., \& Mandel, L. 1987, Physical Review Letters,
Vol. 59, No. 18, 2044

\noindent
Koschmieder, E.L. 1993, Benard Cells and Taylor Vortices (Cambridge:
Cambridge Univ. Press)

\noindent
Landau, L.D., \& Peierls, R. 1931, Z. Phys. 69, 56

\noindent
Loudon, R. 1983, The Quantum Theory of Light (Oxford: Oxford Univ. Press)

\noindent
Mach, E. 1911, History and Root of the Principle of the
Conservation of Energy, P. Jourdain, trans. Chicago: Open Court
(Translation of Mach 1872), p.61

\noindent
Milonni, P.W. 1994, The Quantum Vacuum: An Introduction to Quantum
Electrodynamics (San Diego: Academic Press)

\noindent
Moore, W.J. 1989, Schrodinger: Life and Thought (Cambridge: Cambridge Univ.
Press)

\noindent
Mott, N.F. 1929, Proc. Royal Soc. London, A126, 79

\noindent
Nambu, Y. 1949, Prog. Theor. Phys. 4, 82

\noindent
Namiki, M., Pascazio, S., \& Nakazayo, H. 1997, Decoherence and Quantum
Measurements (New Jersey: World Scientific)

\noindent
Nicolis, G., \& Prigogine, I. 1977, Self-Organization in Nonequilibrium
\newline \indent Systems (New York: Wiley-Intersciences)

\noindent
Popper, K.R. 1935, The Logic of Scientific Discovery (Vienna: Verlag von
Julius Springer), first English edition published in 1959 by
Hutchinson \& Co (London)

\noindent Prigogine, I. 1997, The End of Certainty (Freepress)

\noindent
Sakurai, J.J. 1985, Modern Quantum Mechanics (New York: Addison-Wesley)

\noindent
Schmiedmayer, J., Chapman, M.S., Ekstrom, C.K., Hammond, T.D.,
Kokorowski, D.A., Lenet, A., Rubenstein, R.A., Smith, E.T., \&
Pritchard, D.E. 1997, in Berman ed. Atom Interferometry (San Diego:
Academic Press), p.2

\noindent
Schrodinger, E. 1926a, Die Naturwissenschaften 28, p.664

\noindent
Schrodinger, E. 1926b, Annalen der Physik (4), vol. 79, p.45

\noindent
Schrodinger, E. 1926c, Annalen der Physik (4), vol. 81, p.109

\noindent
Schrodinger, E. 1930, S.B. Preuss. Akad. Wiss., p.296, see
also W.T. Scott 1967,``Erwin Schrodinger: An Introduction to His Writings'',
(Amherst, Univ. Mass. Press), p.64

\noindent
Schrodinger, E. 1935, Naturwissenschaften 23, 807; English translation
by J.D. Trimmer, 1980, Proc. Am. Phil. Soc. 124, 323

\noindent
Schwinger, J. 1948, Phys. Rev. 74, 1439

\noindent
Schwinger, J. 1949a, Phys. Rev. 75, 651

\noindent
Schwinger, J. 1949b, Phys. Rev. 76, 790

\noindent
Scully, M.O., \& Zubairy, M.S. 1997, Quantum Optics (Cambridge:
Cambridge University Press)

\noindent
Von Newmann, J. 1927, Wahrscheinlichkeits-theoretischer
Aufbau der Quantenmechanik, Gottinger Nachrichten I, No. 10,
245

\noindent
Von Newmann, J. 1932, Mathematische Grundlagen der Quantem Mechanik. (Berlin:
Spinger)

\noindent
Weinberg, S. 1995, The Quantum Theory of Fields (Cambridge: Cambridge Univ.
Press)

\noindent
Weinberg, S. 1980, Phys. Lett. 91B, 51

\noindent
Weisskopf, V.F. 1939, Phys. Rev. 56, 72

\noindent
Wheeler, J.A. 1978, in Mathematical Foundations of Quantum Theory, 
A. R. Marlow ed. (New York: Academic Press), p.9

\noindent
Wheeler, J.A., \& Zurek, W.H. 1983, Quantum Theory and Measurement
(Princeton: Princeton Univ. Press)

\noindent
Wigner, E. 1963, Ameri. J. Phys., 31, 6

\noindent
Yang, C.N. 1987, in Schrodinger: Centenary Celebration of a Polymath

\noindent
Yang, C.N., \& Lee, T.D. 1952, Phys. Rev. 87, 404

\noindent
Zee, A. 2003, Quantum Field Theory in a Nutshell (Princeton: Princeton
Univ. Press)

\noindent
Zhang, X. 1996, Astrophysical Journal, 457, 125

\noindent
Zhang, X. 1998, Astrophysical Journal, 499, 98

\noindent
Zhang, X. 1999, Astrophysical Journal, 518, 613

\noindent
Zhang, X. 2003, invited review presented in the 2nd APCTP 
workshop on astrophysics, Formation and Interaction of Galaxies,
ed. H.M. Lee, Journal of Korean Astronomical Society, 36, 223, 2003
                                                                                
\noindent
Zhang, X. 2004,  invited talk presented
in the international conference on Penetrating Bars through
Masks of Cosmic Dust: Hubble Tuning Fork Strikes a New Note, 
eds. D. Block et al., Astrophysics and Space Science, 319, 317

\noindent
Zhang, X. 2005, Naval Research Lab memorandum report, NRL/MR/7210-05-8866

\noindent
Zurek, W.H. 2002, Los Alamos Science, No. 27, p.2

\vfill
\eject

\vfill
\eject

\end{document}